\documentclass[aps,onecolumn,superscriptaddress]{revtex4}

\usepackage{graphicx,epic,eepic,epsfig,amsmath,latexsym,amssymb,verbatim,subfigure,color}

\usepackage{theorem}

\newtheorem{definition}{Definition}

\newtheorem{lemma}[definition]{Lemma}

\newtheorem{theorem}[definition]{Theorem}

\newtheorem{corollary}[definition]{Corollary}

\def\squareforqed{\hbox{\rlap{$\sqcap$}$\sqcup$}}

\def\qed{\ifmmode\squareforqed\else{\unskip\nobreak\hfil

\penalty50\hskip1em\null\nobreak\hfil\squareforqed

\parfillskip=0pt\finalhyphendemerits=0\endgraf}\fi}

\def\endenv{\ifmmode\;\else{\unskip\nobreak\hfil

\penalty50\hskip1em\null\nobreak\hfil\;

\parfillskip=0pt\finalhyphendemerits=0\endgraf}\fi}

\newenvironment{proof}{\noindent \textbf{{Proof~} }}

\newenvironment{example}{\noindent \textbf{{Example~}}}

\mathchardef\ordinarycolon\mathcode`\:

\mathcode`\:=\string"8000

\def\vcentcolon{\mathrel{\mathop\ordinarycolon}}

\begingroup \catcode`\:=\active

  \lowercase{\endgroup

  \let :\vcentcolon

  }

\newcommand{\nc}{\newcommand}

\nc{\rnc}{\renewcommand}

\nc{\beq}{\begin{equation}}

\nc{\eeq}{{\end{equation}}}

\nc{\beqa}{\begin{eqnarray}}

\nc{\eeqa}{\end{eqnarray}}

\nc{\lbar}[1]{\overline{#1}}

\nc{\bra}[1]{\langle#1|}

\nc{\ket}[1]{|#1\rangle}

\nc{\ketbra}[2]{|#1\rangle\!\langle#2|}

\nc{\braket}[2]{\langle#1|#2\rangle}

\nc{\proj}[1]{| #1\rangle\!\langle #1 |}

\nc{\avg}[1]{\langle#1\rangle}

\nc{\Rank}{\operatorname{Rank}}

\nc{\smfrac}[2]{\mbox{$\frac{#1}{#2}$}}

\nc{\Tr}{\operatorname{Tr}}

\nc{\tr}{\operatorname{Tr}}

\nc{\id}{\operatorname{id}}

\nc{\1}{\openone}

\nc{\ox}{\otimes}

\nc{\dg}{\dagger}

\nc{\dn}{\downarrow}

\nc{\cA}{{\cal A}}

\nc{\cB}{{\cal B}}

\nc{\cC}{{\cal C}}

\nc{\cD}{{\cal D}}

\nc{\cE}{{\cal E}}

\nc{\cF}{{\cal F}}

\nc{\cG}{{\cal G}}

\nc{\cH}{{\cal H}}

\nc{\cI}{{\cal I}}

\nc{\cJ}{{\cal J}}

\nc{\cK}{{\cal K}}

\nc{\cL}{{\cal L}}

\nc{\cM}{{\cal M}}

\nc{\cN}{{\cal N}}

\nc{\cO}{{\cal O}}

\nc{\cP}{{\cal P}}

\nc{\cQ}{{\cal Q}}

\nc{\cR}{{\cal R}}

\nc{\cS}{{\cal S}}

\nc{\cT}{{\cal T}}

\nc{\cX}{{\cal X}}

\nc{\cY}{{\cal Y}}

\nc{\cZ}{{\cal Z}}

\nc{\supp}{{\operatorname{supp}}}

\nc{\var}{\operatorname{var}}

\nc{\rar}{\rightarrow}

\nc{\lrar}{\longrightarrow}

\nc{\polylog}{\operatorname{polylog}}

\def\e{\epsilon}

\nc{\RR}{{{\mathbb R}}}

\nc{\CC}{{{\mathbb C}}}

\nc{\FF}{{{\mathbb F}}}

\nc{\NN}{{{\mathbb N}}}

\nc{\ZZ}{{{\mathbb Z}}}

\nc{\PP}{{{\mathbb P}}}

\nc{\QQ}{{{\mathbb Q}}}

\nc{\UU}{{{\mathbb U}}}

\nc{\EE}{{{\mathbb E}}}

\nc{\Icoh}{{I^{\rm coh}}}

\nc{\Qca}{{Q_{\rm ss}}}

\nc{\Qcaa}{{Q^{(1)}_{\rm ss}}}

\nc{\Dcaa}{{D^{(1)}_{{\rm ss}\rightarrow}}}

\nc{\Dca}{{D_{{\rm ss}\rightarrow}}}

\nc{\be}{\begin{equation}}

\nc{\ee}{{\end{equation}}}

\nc{\bea}{\begin{eqnarray}}

\nc{\eea}{\end{eqnarray}}

\nc{\<}{\langle}

\rnc{\>}{\rangle}

\nc{\Hom}[2]{\mbox{Hom}(\CC^{#1},\CC^{#2})}

\nc{\rU}{\mbox{U}}

\def\lpm{ \left(\rule{0pt}{2.1ex}\right. \!}

\def\rpm{ \!\left.\rule{0pt}{2.1ex}\right) }

\begin{document}
\author{Debbie Leung}

\affiliation{Institute for Quantum Computing, University of Waterloo, Waterloo, Ontario, N2L 3G1, Canada} 

\email{wcleung@iqc.ca}

\author{Graeme Smith}

\affiliation{IBM TJ Watson Research Center, 1101 Kitchawan Road, Yorktown Heights, NY 10598, USA}

\email{graemesm@us.ibm.com}

\title{Continuity of quantum channel capacities}

\begin{abstract}
We prove that a broad array of capacities of a quantum channel are
continuous.  That is, two channels that are close with respect to the
diamond norm have correspondingly similar communication capabilities.
We first show that the classical capacity, quantum capacity,
and private classical capacity are continuous,
with the variation on arguments $\e$ apart bounded by
a simple function of $\e$ and the channel's output dimension.
Our main tool is an upper bound of the variation of output entropies
of many copies of two nearby channels given the same initial state; 
the bound is linear in the number of copies.
Our second proof is concerned with the quantum capacities in the
presence of free backward or two-way public classical communication.
These capacities are proved continuous on the interior of the set of
non-zero capacity channels by considering mutual simulation between
similar channels.  

\end{abstract}

\date{\today}

\maketitle

\parskip=2ex

\parindent=0ex

\section{Introduction}

There are several notions of capacity for a noisy quantum
communication channel.  For example, we may be interested in a
channel's capacity for either classical
\cite{Holevo98a,Schumacher97a}, private classical \cite{D03}, or
quantum \cite{Lloyd97,Shor02,D03} communications.  We may have access
to auxiliary resources in addition to the channel, such as
entanglement, one-way classical communication from the sender to
receiver, from the receiver to the sender, or two-way classical
communications.  In all of these situations, there is a sensible
notion of capacity that can be studied.  Except when free auxiliary
entanglement is available, where the problem is effectively solved
\cite{BSST02}, the various capacities of even very simple channels are
unknown.

One property that we would hope for in a capacity is continuity.  From a
practical point of view there will always be a certain amount of
channel uncertainty in real systems.  In this setting, if nearby
channels had dramatically different capacities, the theory of quantum
capacities would be of limited value.  However, from a mathematical
point of view continuity is not at all obvious---very similar channels
can become quite far apart given many copies, and the capacity is
operationally defined in terms of an asymptotic number of channel
uses.  This is not a problem when a single-letter capacity formula is
available, in which case we can reason about the formula directly, but
when only a multi-letter formula is available (or worse, none at all)
the problem of continuity becomes a challenge.

The continuity of channel capacities has been considered before.  For
example, in their study of the quantum erasure channel \cite{BDS97},
Bennett, DiVincenzo, and Smolin implicitly assumed the continuity of
the quantum channel capacity to upper bound the capacity of this
channel.  For the erasure channel, this assumption was rigorously
justified later in \cite{BST98}.  Keyl and Werner explicitly
considered continuity of the quantum channel capacity in \cite{KW02},
where it was shown that the capacity is lower semi-continuous.
Continuity of the Holevo information (whose regularization gives a
multi-letter formula for the classical capacity) was considered in
\cite{Shir06}, where it was shown to be continuous for finite
dimensional outputs and lower semi-continuous in general.

A related set of questions concerns the continuity of entropic
quantities and entanglement measures, which are functions on quantum
states.  For example, Fannes \cite{Fannes73} found a tight bound
on the variation of von Neumann entropy of finite dimensional states.
This was subsequently used by Nielsen to study the continuity of
entanglement of formation \cite{Nielsen98}.  
%
%
As another example, Donald and Horodecki proved the continuity of 
the relative entropy of entanglement \cite{DH99}.
The continuity of
asymptotic (i.e., regularized) entanglement measures was studied by
Vidal in \cite{V02}, which were shown to be continuous in any open set of
distillable states.  More recently, Alicki and Fannes generalized the
continuity result in \cite{AF04} to conditional entropy, and used it to 
prove the continuity of squashed entanglement \cite{CW03}.

In this work we show the continuity of various communication
capacities of quantum channels with finite output dimensions.
For the unassisted capacities for classical, private classical, and
quantum communication, our tool is an inequality controlling the
variation of output entropies of many copies of two nearby channels
given the same initial state.  By careful use of the Alicki-Fannes
inequality \cite{AF04}, this bound is shown to be linear (not
quadratic) in the number of copies.
For the quantum capacity with two-way classical communication, and the
quantum capacity with classical back communication, we also show
continuity within an open set of nonzero quantum capacity channels.
Our results in this setting build on \cite{V02}, whose arguments are
extended from the distillable entanglement of states to the capacity
of channels.

The rest of the paper is organized as follows.  
Section \ref{sec:prelim} contains various definitions, concepts, and
prior results used in this paper.
Our main tool, the inequality controlling the variation of output
entropies of many copies of two nearby channels given the same initial
state is proved in Sec.\ \ref{sec:mainlemma}.
This is used to show our main results, the continuity of the quantum,
classical, and private classical capacity in Sec.\ \ref{sec:cty1way}.
For simplicity throughout most of this paper, we focus on channels with finite dimensional
inputs and outputs, although the results of Sec.\ \ref{sec:cty1way} can easily be seen to apply to channels with
infinite dimensional inputs and finite outputs.  One exception to this focus is in Sec.\ \ref{sec:discty1way}, where we
consider a family of pairs of infinite dimensional channels, parameterized by $n$.  As $n$
increases, each pair has decreasing
distance, but their capacities differ by at least a constant, thereby 
showing that finite output dimension is needed for continuity.  
Continuity for the quantum capacities assisted by backward or two-way
classical communication in the interior of the nonzero capacity region
is proved in Sec.\ \ref{sec:ctyassisted}.  
We make a few concluding remarks in Sec.\ \ref{sec:discussion}.

\section{Preliminaries}
\label{sec:prelim}

In this section, we introduce the concepts, notations, definitions,
and background materials, focusing on finite dimensional quantum
systems.  Notations and discussion in the infinite dimensional case
will be deferred to Sec.\ \ref{sec:discty1way}.

\subsection{Quantum States and Channels}

Let $\cH$ be a complex Hilbert space, and $\cB(\cH)$ be the set of
bounded linear operators taking $\cH$ to itself.  A quantum state is
represented by a positive semidefinite operator $\rho \in \cB(\cH)$
with unit trace.  Except in Sec.\ \ref{sec:discty1way}, we will be
interested in finite-dimensional $\cH$.  A quantum channel $\cN$ that
takes states from $\cH_{\rm in}$ to $\cH_{\rm out}$ is a linear map
from $\cB(\cH_{\rm in})$ to $\cB(\cH_{\rm out})$ that is
trace-preserving and completely-positive.  In particular, when
$\cH=\cH_{\rm in}=\cH_{\rm out}$, we denote by $\cI$ the identity map
from $\cB(\cH)$ to itself.
Recall the definition that $\cN$ is completely-positive if for any 
reference system with associated Hilbert space $\cH_{\rm ref}$, $\cI
\otimes \cN$ maps the positive-semidefinite cone in $\cB(\cH_{\rm ref}
\otimes \cH_{\rm in})$ to that in $\cB(\cH_{\rm ref} \otimes \cH_{\rm
  out})$.  
We also call channels, which are trace-preserving and completely-positive,
``TCP maps.''  They are exactly the physical operations on a state
that are allowed by quantum mechanics.  A quantum system is associated
with a Hilbert space and its set of bounded operators.  We also use the
system name loosely.  For example, we may say that a channel takes
system $A$ to system $B$, or write $\cN:A \rightarrow B$.

We denote the trace, which is a simple example of a TCP map, by
$\tr{[\cdot]}$.  A partial trace on a composite system is simply the
trace operation on one component.  A pure state is a rank one
projector, and is also represented by any vector it projects onto.
For a quantum state $\rho \in \cB(\cH)$, a purification is any pure
state $|\psi\>\<\psi| \in \cB(\cH \otimes \cH')$ such that the partial
trace over $\cH'$ gives $\rho$, and purifications always exist.  Any
channel $\cN$ can be represented as a conjugation by an isometry $U:
\cH_{\rm in} \rightarrow \cH_{\rm out} \ox \cH_{\rm env}$, followed by
a partial trace: $\cN(\rho) = \Tr_{\rm env} U\rho U^\dg$.  

We sometimes add subscripts to the symbols for quantum states and
channels to emphasize what systems they act on, but we may omit these
to avoid cluttering.  However, for multipartite states, the reduced
state on a subset of systems is always subscripted by the subset.  

Throughout this paper, we use a distance measure between states given
by the $1$-norm of their difference: 
\begin{equation}
||\rho-\sigma||_1 = \Tr|\rho-\sigma|
\end{equation}
Half of the above is called the trace distance, the quantum analogue
of the total variation distance in the classical setting.

We use a distance measure between channels (mapping from  
$\cB(\cH_{\rm in})$ to
$\cB(\cH_{\rm out})$) induced by the diamond norm:
\begin{equation}
||\cN_1 - \cN_2 ||_\diamond = \max \{ || (\cN_1 - \cN_2) \otimes \cI
(X)||_1 : X \in \cB(\cH_{\rm in} \otimes \cH_{\rm ref}), ||X||_1 = 1\} \,.
\end{equation}
The maximum can always be attained with $X$ being a pure quantum
state.  Operationally, the diamond norm on the difference between the
two channels characterizes the probability to distinguish them, if one
can prepare an optimal state and feed part of it into the channel.  
The distance measure also has the nice property that, increasing the
dimension of the reference system beyond $\dim(\cH_{\rm in})$ does not
increase the distinguishability.
This gives us control over the trace distance of the output states of
different channels given the same input, and subsequently other
quantities of interest to be defined in the next subsection.  

The diamond norm of a channel is closely related to the family of completely bounded norms (cb-norms), and 
in fact is equal to the usual cb-norm of the adjoint channel as well as a generalized cb-norm of the original channel
(for more on cb-norms and their relation to quantum information, see \cite{Paulsen87,DJKR}).

\subsection{Entropic Quantities}

For a classical random variable $X$ with Prob$(X{\,=\,}x)=p_x$, the Shannon
entropy of $X$ is given by $H(X) = -\sum_x p_x \log p_x$ (or
$H(\{p_x\})$).  If $X$ is binary with probabilities $p,1-p$, $H(X)$ is
written as $H(p)$.  Here and throughout this paper, $\log$ is in base
$2$.

For a quantum system $A$ prepared in state $\rho$, the von Neumann
entropy is written as $S(A)_{\rho}$ or $S(\rho) = -\tr \rho \log \rho=
H(\{ \lambda_k \})$ where $\lambda_k$ is the $k$th eigenvalue of
$\rho$.  Throughout the paper, subscripts showing states on which
entropies and other information theoretic quantities are evaluated are
omitted when there is little risk of confusion.

For two systems $AB$ in state $\rho$, we mention a few measures of
correlation between $A$ and $B$:

$\bullet$ the quantum mutual information is defined as $I(A;B)_\rho
=S(A) + S(B) - S(AB)$ where entropies are evaluated on $\rho$ and its
partial traces.  \\
$\bullet$ The conditional entropy is given by $S(A|B) = S(AB) -
S(B)$. \\
$\bullet$ The coherent information $I^{\rm coh}(A \rangle B)_\rho$ is
given by $S(B)- S(AB) = -S(A|B)$.

The entropy and conditional entropy, viewed as functions of the
underlying states, are both continuous.  The following, particularly
Theorem \ref{Thm:AF}, will be helpful tools for our task of showing
the continuity of capacities.
\begin{theorem}[Fannes Inequality \cite{Fannes73}] 
For any $\rho$ and $\sigma$ with $||\rho-\sigma||_{1} \leq \e$, 
$|S(\rho) - S(\sigma)| \leq \e \log d + H(\e)$. 
\end{theorem}
\begin{theorem}[Alicki-Fannes Inequality \cite{AF04}] \label{Thm:AF}
For any $\rho_{AB}$ and $\sigma_{AB}$ with $||\rho-\sigma||_{1} \leq \e$, 
\begin{equation}
\left|S(A|B)_\rho - S(A|B)_\sigma\right| \leq 4 \e \log d_A + 2 H(\e).
\end{equation}
\end{theorem}

\subsection{Capacities of a quantum channel}

Consider a quantum channel $\cN:A' \rightarrow B$.  The channel $\cN$
has several different capacities for communication.  The following 
quantities will play crucial roles in the various capacities.

$\bullet$ For an input ensemble $\{p_x,\phi_x\}$, let $\omega =
\sum_{x} p_x \proj{x}_X \otimes \cN(\phi_x)$ and 
\begin{equation}
\chi(\cN) := \max_{p_x,\phi_x} I(X;B)_{\omega}
\end{equation}
be the optimal Holevo information \cite{Holevo73} of the output ensemble
(after the channel acts on the input).
%

$\bullet$ For an input state $\rho_{AA'}$, where part of it will be
fed into $\cN$, let
\begin{equation}
I^{\rm coh}(\cN,\rho_{AA'}) = I^{\rm coh}(A\rangle B)_{\cI \ox \cN(\rho_{AA'})}
\end{equation}
be the coherent information generated.
Maximizing over the input gives the coherent information of $\cN$:
\begin{equation}
I^{\rm coh}(\cN) = \max_{\rho_{AA'}} I^{\rm coh}(\cN,\rho_{AA'}) \,.
\end{equation}
We remark that the maximizing state can be chosen to be pure. 

$\bullet$ For an input ensemble $\{p_x,\phi_x\}$, let $\omega =
\sum_{x} p_x \proj{x}_X \ox (U\phi_x U^\dg)_{BE}$, where $U:\cH_{A'}
\rightarrow \cH_B \ox \cH_E$ is an isometric extension of $\cN$.
Then,
\begin{equation}
I^{\rm priv}(\cN) = \max_{p_x,\phi_x} 
                    \left( I(X;B)_{\omega}-I(X;E)_\omega\right) \,,
\end{equation}
where the mutual information is evaluated on the reduced states.  

To give the operational definitions of the different capacities of
$\cN$ for communication, we need to consider $n$ uses of the channel.
We will use shorthands $\cN^n$, $A'^n$, $B^n$, and $E^n$ to stand for
$\cN^{\ox n}$, $A'^{\ox n}$, $B^{\ox n}$, and $E^{\ox n}$.

\begin{definition}Classical Capacity.  We say that a rate $R$ is 
$\e$-classically-achievable if there is an $n_{\e}$ such that for all
  $n \geq n_\e$ there is a classical code $\{ \rho_{k} \in
  A'^n\}_{k=1}^{K_n}$ and a decoding operation $\cD_{n}:B^{n}
  \rightarrow \{\proj{k}\}_{k=1}^{K_n}$ such that $\forall k$, $||
  \cD_{n}(\cN^{n}(\rho_k)) - \proj{k}||_1 \leq \e$ with $\log K_n \geq
  nR$.  A rate is classically-achievable if it is $\e$-classically
  achievable for all $\e>0$.  The classical capacity of $\cN$,
  $C(\cN)$, is the supremum over classically-achievable rates.
\end{definition}

\begin{theorem}{\bf (HSW Theorem \cite{Holevo98a,Schumacher97a})}  
The classical capacity satisfies
\begin{equation}
C(\cN) = \lim_{n\rightarrow \infty}\frac{1}{n} \chi(\cN^n) \,.
\end{equation}
\end{theorem}

\begin{definition}Quantum Capacity.  
We say that a rate $R$ is $\e$-achievable if there is an $n_\e$ such
that for all $n \geq n_\e$ there is a quantum code, $C_n \subset
A'^{n}$ and decoding operation $\cD_n:B^{n}\rightarrow C_n$
such that for all $\psi \in \cB(C_n)$, $||\cD_n(\cN^{n}(\psi)) -
\psi||_1\leq \e$ and $\log \dim \cH_{C_n }\geq nR$.  A rate $R$ is
achievable if it is $\e$-achievable for all $\e>0$.  The quantum
capacity of $\cN$, $Q(\cN)$, is the supremum over achievable rates.
\end{definition}


%

\begin{theorem}{\bf (LSD Theorem \cite{Lloyd97,Shor02,D03})}  
The quantum capacity satisfies
\begin{equation}
Q(\cN) = \lim_{n \rightarrow \infty}\frac{1}{n}I^{\rm coh}(\cN^{n}).
\end{equation}
\end{theorem}

\begin{definition}Private Capacity.  
The private capacity is the capacity of a channel for classical
communication with the added requirement that an adversary with access
to the environment of the channel is ignorant of the communication.
More formally, we say that a rate $R$ is $\e$-privately-achievable if
there is an $n_\e$ such that $\forall n \geq n_\e$ there exists a
classical code $\{ \rho_k \in A'^{n}\}_{k=1}^{K_n}$ with $\log
K_n \geq nR$ and decoding operation $\cD_n:B^{n} \rightarrow
\{\proj{k}\}_{k=1}^{K_n}$ such that for all $k$
\begin{eqnarray}
|| \cD_n(\cN^n(\rho_k)) - \proj{k} \, ||_1 & \leq \e
\\[1ex]
{\rm and} \hspace*{13ex}
|| \rho^{k}_{E^{n}} - \sigma_{E^{n}} ||_1 & \leq \e.
\end{eqnarray}
Here $\rho^k_{E^{n}} = \widehat{\cN}^{n}(\rho_k)$, where
$\widehat{\cN}(\rho) = \tr_B U\rho U^\dg$, with $U:\cH_{A'}
\rightarrow \cH_B \ox \cH_{E}$ an isometric extension of $\cN$, and
$\sigma_{E^{n}}$ is a fixed state on $E^{n}$.  If $R$ is
$\e$-privately-achievable for all $\e>0$, it is called privately
achievable, and the supremum of privately-achievable rates is called
the private capacity.
\end{definition}

\begin{theorem}{\bf (\cite{D03})}
The private capacity satisfies
\begin{equation}
C_p(\cN) = \lim_{n\rightarrow \infty}\frac{1}{n}I^{\rm priv}(\cN^{n}) \,.
\end{equation}
\end{theorem}

The three capacity definitions above are similar in structure, and
differing only in the type of information being sent.  The
corresponding theorems, which give what are called ``regularized
capacity formulas'' also seem to be parallel.  In each case, the
``regularization", as the limit over $n$ is called, prevents us from
evaluating the capacity of a given channel explicitly, or even
numerically.
In the case of the quantum capacity \cite{SS96,SmithSmo0506} and the
private classical capacity \cite{SRS08} it is known for a while that
the regularization cannot be removed in general.
More recently, the regularization in the classical capacity was 
reported to be generally necessary \cite{Hastings-add-08}. 

While very little is known about the capacities above, even less is
known about the capacity of a channel for quantum communication
assisted by two-way classical communication.  
%
%
%
To define this capacity, we introduce the notion of an $n$-use
protocol $\cP_n$, where $n$ denote the number of times the channel
$\cN$ can be used.  Just as in the definition of the unassisted
quantum capacity, we consider a system $C_n$ which holds the quantum
information to be sent.  We use the same symbol to denote Bob's
quantum system which holds the quantum data in his posession at the
end of the protocol.  $\cP_n$ is a composition of the following steps
(in order of being performed): $\cA_0$, $\cM_{\rightarrow 0}$, $\cN$,
$\cB_1$, $\cM_{\leftarrow 1}$, $\cA_1$, $\cM_{\rightarrow 1}$, $\cN$,
$\cB_2$, $\cM_{\leftarrow 2}$, $\cdots$ $\cA_{n{-}1}$,
$\cM_{\rightarrow (n{-}1)}$, $\cN$, $\cB_n$, $\cM_{\leftarrow n}$,
$\cA_{n}$.  Here, each $\cA_i$ is performed by the sender Alice on
$C_n$ and her auxiliary system after the $i$-th channel use,
and each produces an extra system $A'$ as an input to the $(i{+}1)$-th
channel use.  Each $\cM_{\rightarrow i}$ transmits classical
communication from Alice to the receiver Bob.  Each $\cB_i$ is
performed by Bob on his auxiliary system and all $i$ systems cumulated
from the channel uses.  Each generates some classical outcome to be
sent to Alice in the step $\cM_{\leftarrow i}$.  Using the notion of a
protocol, we can now define quantum capacity with two-way classical
assistance.

\begin{definition}Quantum Capacity with two-way classical assistance.

For any $\e>0$ we say that a rate $R$ is $\e$-$2$-way-achievable if
there is an $n_\e$ such that for all $n \geq n_\e$ there is an $n$-use
protocol $\cP_n$ such that for any auxiliary reference system $A$,
$\psi \in C_n \otimes A$, $||\cP_n \otimes \cI (\psi) - \psi||_1\leq
\e$ and $\log \dim \cH_{C_n }\geq nR$.  In other words, $\cP_n$ and
the identity map on the code space are $\e$-close in the diamond norm.
A rate is achievable if it is $\e$-achievable for all $\e>0$.  The
quantum capacity of $\cN$ with two-way classical assistance,
$Q_2(\cN)$, is the supremum over achievable rates.

\end{definition}


\begin{definition}Quantum Capacity with back classical assistance $Q_B(\cN)$.

An $n$-use protocol in this setting is similar to that with two-way
assistance, except that $\cM_{\rightarrow i}$ are omitted.  The rest
of the capacity definition is similar to that of $Q_2(\cN)$.

\end{definition}

Little is known about these assisted capacities.  One proven fact
\cite{BDSW96,HHH00} is that $Q_2(\cN)$ is equal to the {\em
entanglement capacity} of $\cN$ (informally, that is the maximum
amount of near perfect entanglement generated per use of $\cN$,
asymptotically).  Clearly $Q(\cN) \leq Q_B(\cN) \leq Q_2(\cN)$, but
beyond that, almost nothing is known about $Q_B(\cN)$.  For instance,
there is no known analogue of a connection to entanglement capacity.

\section{Continuity of Output Entropy}

\label{sec:mainlemma}


The following theorem is one of our main technical tools.

\begin{theorem} \label{thm:main}
Let $\cN:A'\rightarrow B$ and ${\cM}:A'\rightarrow B$ be quantum
channels and $d_B$ be the finite dimension of $B$.  Let $A$ be an
auxiliary reference system.  If $||\cN - {\cM}||_{\diamond} \leq \e$, 
then, for any state $\phi \in \cB(AA'^{n})$, 
\begin{equation}
\biggl|S\left( (\cI\ox \cN^{n})(\phi)\right) - S\left( (\cI\ox {\cM}^{n})(\phi)\right)\biggr| 
\leq n \left( 4\e\log d_B + 2H(\e)\right).
\end{equation}
\end{theorem}

\begin{proof}
Let 
\begin{equation}
\rho^{k}_{AB^{n}} = \left(\cI_A \ox {\cM}^{\ox k}\ox \cN^{\ox (n-k)}\right)(\phi_{AA'^{n}}) \,.
\end{equation}
In the above, we have explicitly labeled the auxiliary, the input and the 
output systems on the states.  We omit these subscripts from now on.  
Setting $k=0$ and then $n$, we have in particular $\rho^0_{AB^{n}} =
\cI\ox \cN^{n}(\phi)$ and $\rho^{n}_{AB^{n}} = \cI\ox
{\cM}^{n}(\phi)$.  Since $\rho^{k{-}1}$ and $\rho_k$ differs only in the 
$k$th output system, 
\begin{equation} 
\label{eq:eqbydef}
S(AB_1 \dots B_{k-1}B_{k+1}\dots B_n)_{\rho^{k{-}1}} = S(AB_1 \dots B_{k-1}B_{k+1}\dots B_n)_{\rho^{k}}. 
\end{equation}
The quantity we are interested in is
\begin{equation}
\left| S(AB^{n})_{\rho^0} - S(AB^{n})_{\rho^n} \right|, 
\end{equation}
which satisfies
\begin{eqnarray}
\left| S(AB^{n})_{\rho^0} - S(AB^{n})_{\rho^n} \right| &=& 
\left| \; \sum_{k=1}^{n}S(AB^{n})_{\rho^{k{-}1}}-S(AB^{n})_{\rho^{k}}\right|\\
& \leq & \sum_{k=1}^{n}\left| S(AB^{n})_{\rho^{k{-}1}}-S(AB^{n})_{\rho^{k}} \right|.
\end{eqnarray}
Applying Eq.(\ref{eq:eqbydef}) to a single term in this sum, we have
\begin{eqnarray}
\bigl|S(AB^{n})_{\rho^{k{-}1}} \!\!\! & - \! & S(AB^{n})_{\rho^{k}} \bigr| 
\hspace*{70ex}
\nonumber
\\
\nonumber
&= & \bigl|S(AB^n)_{\rho^{k{-}1}} - S(AB_1 \dots B_{k-1}B_{k+1}\dots B_n)_{\rho^{k{-}1}}-S(AB^n)_{\rho^{k}} 
+ S(AB_1 \dots B_{k-1}B_{k+1}\dots B_n)_{\rho^{k}}\bigr|
\\
& = &\bigl| S(B_k|AB_1 \dots B_{k-1}B_{k+1}\dots B_n)_{\rho^{k{-}1}} 
- S(B_k|AB_1 \dots B_{k-1}B_{k+1}\dots B_n)_{\rho^{k}}\bigr|.
\end{eqnarray}
Because $||\cN-{\cM}||_{\diamond} \leq \e$, we also have $||\rho^k -\rho^{k-1}||_1 \leq \e$, 
so by the Alicki-Fannes Inequality, 
\begin{equation}
\bigl| S(B_k|AB_1 \dots B_{k-1}B_{k+1}\dots B_n)_{\rho^k} 
- S(B_k|AB_1 \dots B_{k-1}B_{k+1}\dots B_n)_{\rho^{k{-}1}}\bigr| \leq 4\e \log d_B + 2H(\e).
\end{equation}
As a result, we find
\begin{equation}
\bigl| S(AB^n)_{\rho^0} - S(AB^n)_{\rho^n}\bigr| \leq n(4\e \log d_B + 2H(\e)),
\end{equation}
which completes the proof. $\hfill \square$
\end{proof}

\section{Continuity of Capacities for Channels with finite output dimension}
\label{sec:cty1way}

We now apply Theorem \ref{thm:main} to show the continuity of
$C(\cN)$, $Q(\cN)$, and $C_p(\cN)$.  Each of these capacities has the
form $F(\cN) = \lim_{n \rightarrow \infty} \frac{1}{n} \max_{P^{(n)}}
f_n(\cN^n, P^{(n)})$ for some appropriate family of function $\{f_n\}$ 
and parameters $P^{(n)}$ to be optimized over.
We make repeated use of the following Lemma.  
\begin{lemma}
\label{lem:oplem}
If $F(\cN) = \lim_{n \rightarrow \infty} \frac{1}{n} \sup_{P^{(n)}}
f_n(\cN^n, P^{(n)})$ and $\forall n$, $\forall P^{(n)}$, 
$|f_n(\cN^n, P^{(n)})-f_n(\cM^n, P^{(n)})| \leq n c$, then 
$|F(\cN) - F(\cM)| \leq c$.  
\end{lemma}
\begin{proof}
Let $\e>0$ be arbitrary.  
Let $f_n(\cN^n) = \sup_{P^{(n)}} f_n(\cN^n, P^{(n)})$.  
Suppose $f_n(\cN^n)$ and $f_n(\cM^n)$ are $\epsilon$-close to optimal
at $P_1^{(n)}$ and $P_2^{(n)}$.
Then, 
\bea
f_n(\cN^n) - \e < f_n(\cN^n, P_1^{(n)}) \leq f_n(\cM^n, P_1^{(n)}) +
nc \leq f_n(\cM^n) + nc \\
f_n(\cM^n) - \e < f_n(\cM^n, P_2^{(n)}) \leq f_n(\cN^n, P_2^{(n)}) +
nc \leq f_n(\cN^n) + nc 
\eea
Thus, $\forall \e>0, \forall n$, 
$|f_n(\cN^n) - f_n(\cM^n)| \leq nc+\e$.  Taking 
limits $\e \rightarrow 0, n \rightarrow \infty$, 
$|F(\cN) - F(\cM)| \leq c$.  

\end{proof}

Note that in particular, Lemma \ref{lem:oplem} holds with $\sup$
replaced by $\max$, as needed in the following corollaries.

\begin{corollary}
The classical capacity of a quantum channel with finite-dimensional
output is continuous.  Quantitatively, if $\cN, \cM:A' \rightarrow B$
where the dimension of $B$ is $d_B$ and $||\cN - \cM||_{\diamond} \leq
\e$, then
\begin{equation}
|C(\cN)-C(\cM)| \leq 8 \e \log d_B + 4 H(\e).
\end{equation}
\end{corollary}

\begin{proof}
From the HSW theorem 
\begin{equation}
C(\cN) = \lim_{n\rightarrow \infty}\frac{1}{n} \chi(\cN^n) = 
\lim_{n\rightarrow \infty}\frac{1}{n} \max_{p_x,
  \phi_x^{(n)}} I(X;B^n)_{\omega^{(n)}},
\end{equation}
where $\omega^{(n)} = \sum_{x}p_x \proj{x}_X \otimes \cN^{\ox
  n}(\phi_x^{(n)})$.  For any $\cN:A' \rightarrow B$ and
$\cM:A' \rightarrow B$ with $||\cN - \cM||_\diamond \leq \e$, and for
fixed $n$ and $\{p_x, \phi_x^{(n)}\}$, letting $\omega = \sum_x p_x
\proj{x}_X \otimes \cN^{n}(\phi_x^{(n)})$ and $\tilde{\omega} =
\sum_x p_x \proj{x}_X \ox \cM^{n}(\phi_x^{(n)})$, we have
\begin{eqnarray}
\left|I(X;B^n)_{\omega} - I(X; B^n)_{\tilde{\omega}} \right|
& = & \left| S(B^n)_{\omega}- S(B^nX)_{\omega}-S(B^n)_{\tilde{\omega}} +
S(B^nX)_{\tilde{\omega}} \right|
\\
& \leq & \left| S(B^n)_{\omega}-S(B^n)_{\tilde{\omega}}\right| +
\left|S(B^nX)_{\tilde{\omega}} - S(B^nX)_{\omega}\right|
\\
& \leq & 2n \left( 4\e \log d_B + 2H(\e) \right).
\end{eqnarray}
Applying Lemma \ref{lem:oplem} gives the desired result $|C(\cN)-
C(\cM)| \leq 8\e \log d_B + 4H(\e)$.  $\hfill \square$
\end{proof}

\begin{corollary}

The quantum capacity of a quantum channel with finite dimensional
output is continuous.  Quantitatively, if $\cN, \cM:A' \rightarrow B$
where the dimension of $B$ is $d_B$ and $||\cN - \cM||_{\diamond} \leq
\e$, then
\begin{equation}
|Q(\cN)-Q(\cM)| \leq 8 \e \log d_B + 4 H(\e).
\end{equation}
\end{corollary}

\begin{proof}
From the LSD Theorem, 
\begin{equation}
Q(\cN) = \lim_{n \rightarrow \infty}\frac{1}{n}I^{\rm coh}(\cN^{n}) 
= \lim_{n \rightarrow \infty}\frac{1}{n}
  \max_{\rho_{AA'^n}} I^{\rm coh}(\cN^n,\rho_{AA'^n}) \,.
\end{equation}
Let $\omega_{AB^n} = \cI \ox \cN^{n}(\rho_{AA'^n})$ and
$\tilde{\omega}_{AB^n} = \cI \ox \cM^{n}(\rho_{AA'^n})$.  Consider the
difference of coherent informations
\begin{eqnarray}
\left|I^{\rm coh}(\cN^{n},\rho_{AA'^n}) - I^{\rm
coh}(\cM^{n},\rho_{AA'^n})\right|
& = & \left| \; S(B^n)_{\omega} - S(AB^n)_{\omega} -
S(B^n)_{\tilde{\omega}} + S(AB^n)_{\tilde{\omega}} \; \right|
\\ & \leq & \left|S(B^n)_{\omega} - S(B^n)_{\tilde{\omega}} \right| +
            \left|S(AB^n)_{\omega} - S(AB^n)_{\tilde{\omega}}\right|
\\ & \leq & 2n\left( 4\e\log d_B + 2 H(\e)\right).
\end{eqnarray}
Applying Lemma \ref{lem:oplem} gives the result.  $\hfill \square$
\end{proof}

\begin{corollary}
The private classical capacity of a quantum channel with
finite-dimensional output is continuous.  Quantitatively, if $\cN,
\cM:A' \rightarrow B$ where the dimension of $B$ is $d_B$ and 
$||\cN - \cM||_{\diamond} \leq \e$, then
\begin{equation}
|C_p(\cN)-C_p(\cM)| \leq 16 \e \log d_B + 8 H(\e).
\end{equation}
\end{corollary}

\begin{proof}
Let $U$ and $W$ be the isometric extensions for $\cN$ and $\cM$ respectively.  
\begin{equation}
C_p(\cN) = \lim_{n\rightarrow \infty}\frac{1}{n}I^{\rm priv}(\cN^{n})
= \lim_{n\rightarrow \infty} \frac{1}{n} \max_{p_x,\phi_x} \left(
I(X;B^n)_{\omega} -I(X;E^n)_{\omega}\right),
\end{equation}
where $\phi_x$ lives in $A'^n$, $\omega_{X B^n E^n} = \sum_{x} p_x
|x\>\<x| \ox U \phi_x U^\dg$ and $\ket{\omega}_{X B^n E^n G}$ purifies
it.  
Then, 
\begin{eqnarray}
& & 
I(X;B^n)_{\omega} - I(X;E^n)_{\omega}
\\
& = & [S(B^n)-S(B^n X)]_{\omega} 
    - [S(E^n)-S(E^n X)]_{\omega} 
\\
& = & [S(B^n)-S(B^n X)]_{\omega} 
    - [S(X B^n G)-S(B^n G)]_{\proj{\omega}}
\label{eq:tmp}
\end{eqnarray}
Similarly, define $\tilde{\omega}_{X B^n E^n} = \sum_{x} p_x |x\>\<x|
\ox W \phi_x W^\dg$ for $\cM$.  
Switching from Eq.~(\ref{eq:tmp}) to that defined by $\tilde{\omega}$,
the difference can be bounded by applying Theorem \ref{thm:main} to
each of the four terms followed by Lemma \ref{lem:oplem}, giving the
stated result.  $\hfill \square$
\end{proof}

\section{Discontinuity of capacities with infinite output dimension}
\label{sec:discty1way}

In this section we provide simple examples to show that the classical
and quantum capacities of channels with infinite output dimensions are
not generally continuous.  An earlier demonstration of the
discontinuity of the classical capacity for infinite dimensional
quantum channel was given by Shirokov \cite{Shirokov06b}.

For an infinite dimensional complex Hilbert space $\cH$ with bounded
linear operators $\cB(\cH)$, the space of all trace class operators
(subset of $\cB(\cH)$ with finite trace) is denoted
$\mathfrak{T}(\cH)$, and its positive semidefinite subset is denoted
$\mathfrak{T}_{+}(\cH)$.  A quantum state is an element of
$\mathfrak{T}_{+}(\cH)$ with unit trace.  A quantum channel $\cN$ from
$\cH_{\rm in}$ to $\cH_{\rm out}$ is a linear map from
$\mathfrak{T}(\cH_{\rm in})$ to $\mathfrak{T}(\cH_{\rm out})$ that is
trace-preserving and completely-positive.  

\subsection{Classical Capacity}


\begin{example}
Let $\cH = {\rm Span}\{\ket{i}\}_{i=0}^\infty$, and $\cH_+ =
{\rm Span}\{\ket{i}\}_{i=1}^\infty$.  Consider the channels $\cN$ and
$\cM_n:\mathfrak{T}(\cH_+)\rightarrow \mathfrak{T}(\cH)$ with
\begin{equation}
\cN(\ket{i}\bra{j}) = \tr(|i\>\<j|) ~ \proj{0}
\end{equation}
and
\begin{equation}
\cM_n = \lpm 1 - \frac{1}{\log n} \rpm \cN + \frac{1}{\log n}\id_n \,,
\end{equation}
where
\begin{eqnarray}
\id_n(\ket{i}\bra{j}) &=& \ket{i}\bra{j} \hspace*{13ex} 
 {\rm for} \ 1 \leq i,j\leq n\\
 & = & \tr(|i\>\<j|) ~ \proj{0} \ \ \ \ {\rm otherwise}.
\end{eqnarray}
First of all, we have $C(\cN) = 0$, since $\cN$ maps every state to
$\proj{0}$.  
%
%
%
As for the capacity of $\cM_n$, an easy lower bound can be obtained by
using the codewords $|k\>\<k|$ for $k=1,\cdots,n$, turning $\cM_n$ to
a classical erasure channel in $n$-dimensions, with erasure
probability $p_{\rm e} = 1-\frac{1}{\log n}$.  The capacity of the
latter is known\cite{BDS97} to be $(1-p_{\rm e}) \log n = 1$.  Thus, 
\begin{eqnarray}
C(\cM_n) \geq 1.
\end{eqnarray}
However, 
\begin{eqnarray}
|| \cN - \cM_n ||_\diamond &=& \left|\left| \frac{1}{\log n}(\cN - \id_n)
\right|\right|_\diamond\\ & = & \frac{1}{\log n}|| \cN - \id_n||_\diamond
\leq \frac{2}{\log n}.
\end{eqnarray}$\hfill \square$
\end{example}

%

%

\subsection{Quantum Capacity}

\begin{example}
Now let $\cN:\mathfrak{T}(\cH_+)\rightarrow \mathfrak{T}(\cH)$ be defined by
\begin{equation}
\cN(\rho) = \frac{1}{2} \tr(\rho) ~ \proj{0} + \frac{1}{2}\rho.
\end{equation}
That is, $\cN$ is a $50\%$ erasure channel, so that $Q(\cN) = 0$.
Let 
\begin{equation}
\cM_n = \left( 1  -\frac{1}{\log n}\right)\cN + \frac{1}{\log n}\id_n \,.
\end{equation}
A lower bound of the quantum capacity can be obtained by restricting
each input to the span of $\{|i\>\}_{i=1,\cdots,n}$, so that $\cM_n$ is
effectively a quantum erasure channel with $n$-dimensional inputs and with
erasure probability $p_{\rm e} = \frac{1}{2}-\frac{1}{2\log n}$.  
This quantum erasure channel has capacity\cite{BDS97} $(1-2p_{\rm e}) \log n = 1$. 
Therefore, 
\begin{equation}
Q(\cM_n) \geq 1.
\end{equation}
As before, we have $|| \cN - \cM_n||_\diamond \leq \frac{2}{\log n}$,
so that $Q$ is also discontinuous.  $\hfill \square$
\end{example}

\section{Two-way capacity and capacity with back communication}

\label{sec:ctyassisted}

For a general channel, these capacities are not known to have a closed
form expression.  In this setting, an argument similar to that for
continuity of asymptotic entanglement measures in \cite{V02} can be
used for the interior of the nonzero capacity region.  $Q_2$ and $Q_B$
differ in the definition of the $n$-use protocol, and we will see that
this difference does not affect the argument, and we only talk about
$Q_2$ for clarity.

For any metric chosen for the space of channels,
continuity of $Q_2$ at $\cN$ can be stated as $\forall \e>0$, $\exists
\delta>0$ such that $\forall \cN' \in B(\cN,\delta)$, $|Q_2(\cN') -
Q_2(\cN) | \leq \e$, where $B(\cN,\delta)$ is an open ball of radius
$\delta$ centered at $\cN$.  (Similarly for $Q_B$).

We consider the set of channels taking $d_{\rm in}$ to $d_{\rm out}$
dimensions.

\subsection{Interior of $\{Q_2(\cN)>0\}$}

Let us denote the interior of $\{Q_2(\cN)>0\}$ by $\cQ_2^+$. 
Suppose $\cN \in \cQ_2^+$.  Using the definition of continuity stated
above, we will derive $\delta$ as a function of $\e$ and other
relevant parameters, so that $\forall \e>0$, $\exists \delta>0$ such
that $\forall \cN' \in B(\cN,\delta)$, $|Q_2(\cN') - Q_2(\cN) | \leq
\e$.

First, consider $B(\cN,\Delta)$ where $\Delta$ is small enough to
ensure $B(\cN,\Delta) \subset \cQ_2^+$ (i.e., $Q_2>0$ on the entire
$B(\cN,\Delta)$).  Second, for every $\cM$ on the boundary of
$B(\cN,\Delta)$, we specify two other channels $\cM_1$ and $\cM_2$
so that:
\begin{eqnarray} 
\cM = p_1 \cM_1 + (1-p_1) \cN \,, \label{eq:1} \\
\cN = p_2 \cM_2 + (1-p_2) \cM \,, \label{eq:2} 
\end{eqnarray} 
for some $p_1,p_2 \in [0,1]$. 
$\cM_1,\cM_2$ need not be in $\cQ_2^+$ but have to be TCP maps.  
Such $\cM_1,\cM_2$ always exists (for example, we can take them 
to be $\cM$ and its antipodal point on $B(\cN,\Delta)$ respectively).
We take $\cM_1,\cM_2$ to be on the boundary of the set of channels, as
far from $\cN$, $\cM$ as possible to minimize $p_1,p_2$.

The concepts involved in the proof are summarized in the following 
diagram: 
\begin{center}
\includegraphics[scale=1.0]{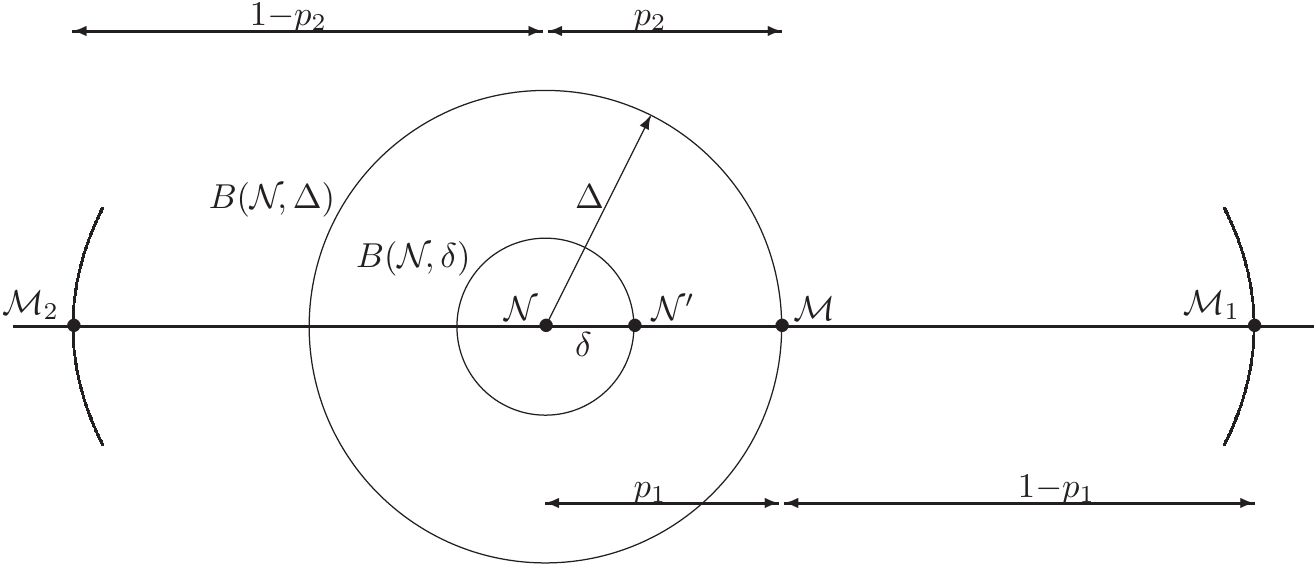}\\
\end{center}

We show how to simulate $\cM$ by $\cN$, from which we derive an upper
bound on $Q_2(\cM)$, Eq.~(\ref{eq:bound1}), in terms of $Q_2(\cN)$.  A
less $\e$-$\delta$-loaded, more concise, and slightly more heuristic
derivation in terms of resource inequalities \cite{DHW03} is given in
\cite{RI}.

(1) We start from the definition of $Q_2(\cN)$.  Consider any $R_1 <
Q_2(\cN)$, with $\delta_1 > 0$ such that $R_1 = Q_2(\cN) - \delta_1$.  For
any $\e > 0$, $\exists n_\e$ such that $\forall n_1 \geq n_\e$, there
is a protocol $\cP_{n_1}$ with $n_1$ uses of $\cN$ and $2$-way
classical communication that simulates the identity map on an $2^{n_1
  R_1}$-dimensional system $\e$-close in diamond norm.

(2) Any channel can be trivially (and inefficiently) simulated by 
either one of the two following methods: Alice sends the input
noiselessly to Bob who then locally applies the channel, or Alice
applies the channel on the input and sends the resulting state to Bob
via the noiseless channel.  Thus, $\log d$ noiseless qubit channels are 
sufficient for simulating any channel where $d = \min(d_{\rm
  in},d_{\rm out})$, in an exact and $1$-shot manner.

(3) Using the assisting classical communication (only one of the
forward or backward direction suffices), Alice and Bob can agree on
$n$ biased coins (with probabilities of the two outcomes being
$p_1,1{-}p_1$) and apply the channel $\cN$ or $\cM_1$ accordingly.
Due to the Chernoff bound, $\forall \delta_{\rm Ch}, \exists n_{\rm
  Ch}$ such that the probability is less than $\e$ that it requires
more than $n (p_1+\delta_{\rm Ch})$ uses of $\cM_1$ or more than $n
(1-p_1+\delta_{\rm Ch})$ uses of $\cN$.  In this unlikely event, Alice
and Bob just run an inaccurate simulation.

We now put these $3$ steps together.  
Let $n_1 = n \, (Q_2(\cN)-\delta_1)^{{-}1} \, (p_1+\delta_{\rm Ch})
\log d$.
We use an $n_1$-use protocol of $\cN$ to simulate $n (p_1+\delta_{\rm
  Ch}) \log d$ identity channels (it will be $\e$-close in diamond
norm if $n_1 \geq n_\e$) which in turns simulates
$n (p_1+\delta_{\rm Ch})$ uses of $\cM_1$ with the same precision.
In addition to the above, we use the coin tosses and $n
(1-p_1+\delta_{\rm Ch})$ direct uses of $\cN$ to simulate $n$ uses of $\cM$.
This simulation is $\e$-close unless an atypical outcome of the coin tosses 
occurs.  
If $n$ is large enough, then $n_1 \geq n_\e$ and $n \geq n_{\rm Ch}$,
the simulation is $2\e$-close in diamond norm.  This takes a total of
$n (1-p_1+\delta_{\rm Ch}) + n \, (Q_2(\cN)-\delta_1)^{{-}1} \, 
(p_1+\delta_{\rm Ch}) \log d$ uses of $\cN$.

Now, $\forall \delta_2 >0$, $R_2=Q_2(\cM)-\delta_2$, $\exists
m_\epsilon$ such that $\forall n \geq m_{\e}$, there is a protocol
with $n$ uses of $\cM$ that simulates the identity map on $2^{nR_2}$
dimensions $\epsilon$-close in diamond norm.  Substitute these $n$
uses of $\cM$ by the $2\e$-close simulation above.  We have an
$3\e$-close simulation of the $2^{nR_2}$-dim identity map with $n
(1-p_1+\delta_{\rm Ch}) + n/[Q_2(\cN)-\delta_1] \times
(p_1+\delta_{\rm Ch}) \log d$ uses of $\cN$.  Letting $\e$,
$\delta_1$, $\delta_2$, and $\delta_{\rm Ch} \rightarrow 0$, we have
\begin{equation} 
\left[ p_1 \frac{\log d}{Q_2(\cN)} + (1-p_1) \right] Q_2(\cN)
     \geq Q_2(\cM)
\label{eq:bound1}
\end{equation}

Running the same argument with $\cN$, $\cM$ reversed and using
Eq.~(\ref{eq:2}) instead, we have 
\begin{equation} 
\left[ p_2 \frac{\log d}{Q_2(\cM)} + (1-p_2) \right] Q_2(\cM) \geq 
Q_2(\cN)
\label{eq:bound2}
\end{equation}
Together, 
\begin{equation} 
|Q_2(\cN) - Q_2(\cM)| \leq \min [ p_1 (\log d -
Q_2(\cN)), p_2 (\log d - Q_2(\cM))] \,.
\label{eq:bound3}
\end{equation}

We now consider $\cN'$ which is colinear with $\cN$ and $\cM$, and is
on the boundary of $B(\cN,\delta)$.  
We can run the same argument with $\cN'$ in place of $\cM$ but with
the same $\cM_1$, $\cM_2$.  
Here, $\cN' = \frac{\delta}{\Delta} \cM + (1-\frac{\delta}{\Delta}) \cN$.  
Eliminating $\cM$ from Eqs.~(\ref{eq:1}) and (\ref{eq:2}), one can
verify that the parameters change as 
\begin{eqnarray}
p_1 & \rightarrow & q_1 = p_1 \frac{\delta}{\Delta} \\ p_2 &
\rightarrow & q_2 = p_2 \frac{\delta}{\Delta} \cdot
\frac{1}{\frac{\delta}{\Delta} p_2 + (1-p_2)} \leq p_2
\frac{\delta}{\Delta} \frac{1}{1-p_2} \leq 2 p_2 \frac{\delta}{\Delta} \,.
\end{eqnarray}
In the last inequality, we use the fact that $p_2 \leq 1/2$ by
construction.
Using Eq.~(\ref{eq:bound3}) for $\cN'$ and substituting $p_1$,
$p_2$ by $q_1$, $q_1$, and for $\delta \leq \frac{\Delta \e}{2 \log d}$,
$$|Q_2(\cN) - Q_2(\cN')| \leq \min [ q_1 (\log d - Q_2(\cN)), q_2
  (\log d - Q_2(\cN'))] \leq \e.$$
Note that $\delta$ depends on $\cN' \in B(\cN,\delta)$ via the
dependence of $\cM'$ and $\Delta$ on $\cN'$.

The continuity bound is not as tight as those derived for the
unassisted capacities, but it has the merit of being independent of
the metric used for the channels.

The same argument holds for continuity of $Q_B$ in the interior of
$Q_B(\cN)>0$ with the only modification in the definition of an 
$n$-use protocol.

\subsection{$Q_B$ of Erasure Channel}

The erasure channel of erasure probability $p$ acts on qubit states as
follows: $\cE_p(\rho) = (1-p) \rho + p |2\>\<2|$, where $|2\>$ can be
view as an error symbol.  $Q_2(\cE_p) = 1-p$ but an expression for
$Q_B(\cE_p)$ is unknown, though it is known to be positive for $p<1$.

%
%

Instead of the continuity of $Q_2$ or $Q_B$ at $\cE_p$, we can ask if
these capacities are continuous as a function of $p$.  In other word,
we are considering the restriction of these functions to the $1$-parameter
family of channels $\cE_p$.

In this restricted domain, $Q_2(\cE_p) = 1-p$ is clearly continuous.
For $Q_B(\cE_p)$, the previous proof now holds on the restricted
domain for $p<1$.  For the point $p=1$, continuity still holds because 
$Q_B(\cE_p) \leq Q_2(\cE_p) = 1-p$ which is vanishing (converging 
towards $Q_B(\cE_1)$) as $p \rightarrow 1$.

\section{Discussion}

\label{sec:discussion}

We have shown that many of the communication capacities of a quantum
channel are continuous.  For unassisted capacities, such as private,
quantum, and classical capacities we proved continuity using Theorem
\ref{thm:main}.  In these cases, the capacities are near-Lipschitz
when the distance between the channels is no less than the inverse of
the single use output dimension.  We obtained explicit bounds on the
effective Lipschitz constants, typically finding variations of order
$\e\log d$ for channels that are distance $\e$ apart.  For the more
involved case of two-way capacity, we have shown continuity of $Q_2$
on the interior of $\{Q(\cN)>0\}$, and similarly for $Q_B$ by making
use of an argument of Vidal\cite{V02}.

In general, application of Theorem \ref{thm:main} will give continuity
any time a regularized capacity formula is available.
In particular, it can easily be used to show the continuity of the
capacity region of multi-user channels such as the multiple access
channel \cite{YDH05} and broadcast channels \cite{YHD06,DH06}.

\section*{Acknowledgements}
We are grateful to Aram Harrow for discussions about continuity of the
two-way and back-assisted capacities, and John Smolin for suggesting
the example of discontinuity for the classical capacity of
infinite-dimensional channels.  We thank Bill Rosgen for a careful
reading and many helpful corrections on an earlier version of the
manuscript.  DL was supported by CRC, CFI, ORF, NSERC, CIFAR, MITACS,
ARO, and QuantumWorks.

\bibliographystyle{apsrev}


\begin{thebibliography}{31}
\bibitem[{RI()}]{RI}
\bibinfo{note}{We describe the proof in the language of resource
  inequalities\cite{DHW03}. Each resource inequality (RI) $S_1 + S_2 \cdots
  \geq R_1 + R_2 \cdots$ represents the fact that $n$ units of LHS resources
  can be used to simulate $n$ units of the RHS resources for asymptotically
  large $n$ and with sufficient accuracy (say, in diamond norm of the
  operations involved in the RHS). If the simulation is sufficiently good,
  manipulating RIs as though they are usual algebraic inequalities can often be
  justified. In particular, the following can be rigorously proved in in many
  situations.\\ (1) Multiplication by a positive scalar on both sides is
  allowed. \\ (2) Inequalities can be summed. \\ (3) The inequalities are
  transitive. \\ (4) Substitution that preserves the inequalities is allowed.
  Operationally, this requires that simulations are accuracy enough to be
  composable, so that recursive/concatenated simulation is possible. In a
  related manner, cancellation (or subtraction) is frequently possible. \\[2ex]
  Now, we give an argument for Eq.~(\ref{eq:bound1}). By the definition of
  two-way assisted channel capacity, \begin{equation} \cN + \infty \; {\rm
  CC}_{\leftrightarrow} \geq Q_2(\cN) \; \cI \,, \end{equation} where $\infty
  \; {\rm CC}_{\leftrightarrow}$ denotes the assistance. Using $\log d$ qubit
  noiseless channels to simulate $\cM_1$ (see main text), \begin{equation} \log
  d \; \cI \geq \cM_1 \,. \end{equation} Together \begin{equation}
  \label{eq:ineq1} \cN + \infty \; {\rm CC}_{\leftrightarrow} \geq Q_2(\cN) \;
  \cI \geq \frac{Q_2(\cN)}{\log d} \; \cM_1 \,. \end{equation} It means that we
  can use $n$ copies of $\cN$ to transmit $n \frac{Q_2(N)}{\log d}$ inputs to
  the receiver, who then applies copies of $\cM_1$ locally thereby giving a
  protocol to simulate $\cM_1$ using $\cN$. Equation (\ref{eq:1}) means that
  \begin{equation} \label{eq:ineq2} p_1 \cM_1 + (1-p_1) \cN + \infty \; {\rm
  CC}_{\leftarrow} \geq \cM \,. \end{equation} Using free back communication to
  generate $n$ biased coins (see main text) \begin{equation} \left[ p_1
  \frac{\log d}{Q_2(\cN)} + (1-p_1) \right] \cN + \infty \; {\rm
  CC}_{\leftrightarrow} \geq \cM \,. \end{equation} This implies
  \begin{equation} \left[ p_1 \frac{\log d}{Q_2(\cN)} + (1-p_1) \right]
  Q_2(\cN) \geq Q_2(\cM) \end{equation} as claimed.}
\end{thebibliography}

\end{document}